\documentclass[a4paper,twocolumn,11pt,unpublished]{quantumarticle}
\pdfoutput=1
\usepackage[utf8]{inputenc}
\usepackage[english]{babel}
\usepackage[T1]{fontenc}
\usepackage{amsmath}
\usepackage{hyperref}

\usepackage{graphicx}
\usepackage{subfigure}
\usepackage{epstopdf}
\usepackage{graphicx}

\usepackage{amsfonts}
\usepackage{amssymb, color, bm}

\usepackage{currvita}

\usepackage{algorithmicx}
\usepackage[noend]{algpseudocode}
\usepackage{algorithm,setspace}

\usepackage{physics}

\usepackage{multicol}
\usepackage{lipsum}
\usepackage{mwe}

\DeclareMathOperator{\E}{\mathbb{E}}

\usepackage{tikz}
\usepackage{lipsum}

\begin{document}

\title{Variational Quantum Integrated  Sensing and Communication}

\author{Ivana Nikoloska}
\affiliation{Signal Processing Systems Group,  Department of Electrical Engineering, Eindhoven University of Technology, Eindhoven, 5612 AP, The Netherlands}
\email{i.nikoloska@tue.nl}
\author{Osvaldo Simeone}
\affiliation{KCLIP lab, Center for Intelligent Information Processing Systems, Department of Engineering, King's College London, Strand, London, WC2R 2LS, United Kingdom}

\maketitle

\begin{abstract}
The integration of sensing and communication  functionalities within a common system is one of the main innovation drivers for  next-generation  networks. In this paper, we introduce an quantum integrated  sensing and communication  (QISAC) protocol that leverages entanglement in quantum carriers of information to enable  both superdense coding and  quantum sensing.  The proposed approach adaptively optimizes  encoding and quantum measurement  via variational circuit learning, while employing classical machine learning-based decoders and estimators to process the measurement outcomes. Numerical results for qudit systems demonstrate that the proposed QISAC protocol can achieve a flexible trade-off between classical communication rate and accuracy of parameter estimation.
\end{abstract}

\section{Introduction }
\noindent \textbf{Context and Motivation:} Wireless networks that jointly perform sensing and communication  have been extensively studied in recent years as a key driver for the development of 6G systems \cite{jeong2014beamforming,Liu2021ISAC,Zhang2021JCR}. By sharing spectrum and hardware resources, such \textit{integrated sensing and communication} (ISAC) systems can improve spectral efficiency and overall functionality, while unlocking new use cases for wireless systems  \cite{Zhang2021JCR}. With the advent of quantum communication technologies and with work under way to define a Quantum Internet \cite{cacciapuoti2019quantum,popovski20251q}, it is natural to consider extensions of ISAC to the quantum domain.

\begin{figure*}
    \centering
    \includegraphics[width = 0.75\textwidth]{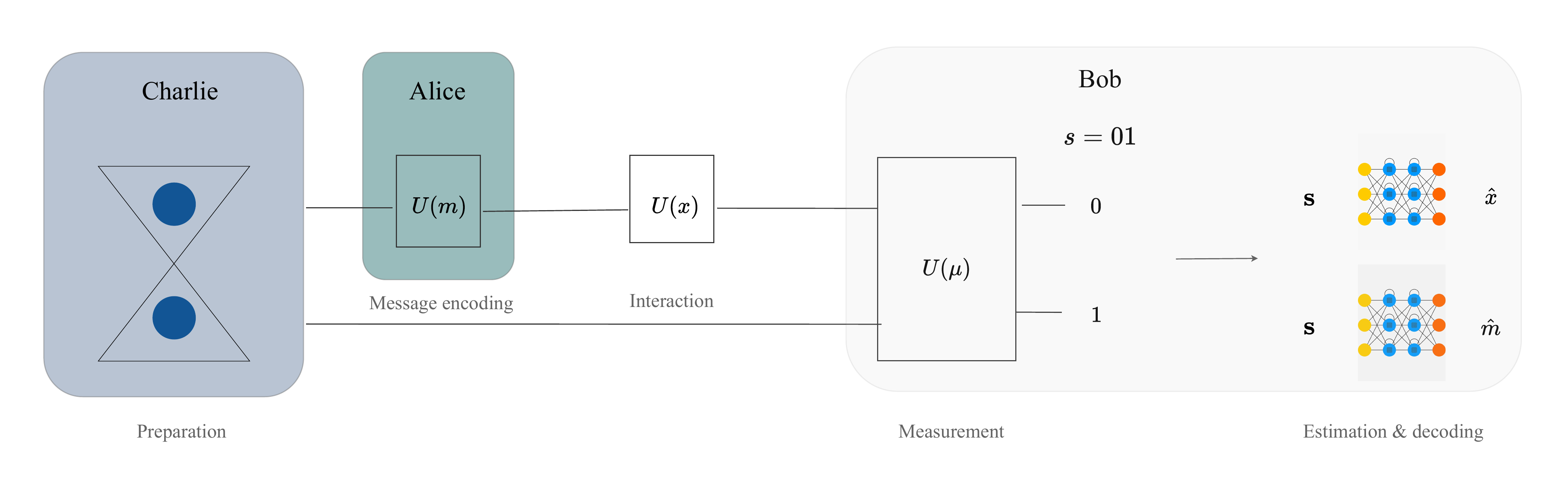}
    \caption{Proposed QISAC protocol: Charlie generates a probe quantum state consisting of two qudits, and distributes the two qudits to Alice and Bob. Alice encodes her message $m$ via a unitary transformation $U(m)$ following superdense coding, and her qudit is sent via the channel, interacting with the parameter of interest, $x$. Bob receives the qudit sent by Alice, applies a decoding unitary $U(\mu)$, and measures the qudit obtain a measurement sample $s$. The protocol uses a classical decoder and estimator which produce the decoded message $\hat{m}$ and the estimate $\hat{x}$, respectively, using the measurement outcome.} 
    \label{fig:sensing_scheme_indirect} 
\end{figure*}

Quantum communication and quantum metrology are two of the most compelling applications of quantum information science, and they  are both realized through the transmission of quantum states \cite{Degen2017}, benefiting from non-classical resources such as entanglement. In particular, entangled qubits enable reliable transmission of classical information at rates beyond classical limits as in superdense coding \cite{Harrow2004}, while the use of entanglement in sensing can achieve precision beyond classical bounds, approaching the Heisenberg limit \cite{Degen2017,Yu2022}. 

 Recently, preliminary studies have begun to explore \emph{quantum integrated sensing and communication } (QISAC), the quantum analog of ISAC.  In particular, reference \cite{Liu2024} proposed an QISAC protocol that achieves quantum sensing under the Heisenberg limit, while simultaneously enabling quantum-secure direct communication via entanglement. The paper \cite{erdougan2025joint} studied the allocation of entanglement resources over bosonic channels, theoretically characterizing the communication-sensing trade-off. Other recent works have further advanced QISAC in optical and bosonic channels \cite{Gong2025,Munar-Vallespir2024Joint}.

Beyond these theoretical insights, it is important to study also practical strategies for QISAC on near-term quantum hardware. Variational quantum algorithms (VQAs) have emerged as a promising approach to navigate the complexity of quantum optimization in the NISQ era \cite{schuld2021machine,simeone2022introduction}. In particular, hybrid quantum-classical optimization of parameterized quantum circuits has been applied to quantum sensing problems \cite{MacLellan2024,Ulum2024,NikoloskaConformal2025}. Notably, reference \cite{MacLellan2024} demonstrated an end-to-end variational quantum sensing method that jointly optimizes state preparation and measurement for a sensing task; the work  \cite{Ulum2024} developed a variational framework for anonymous quantum sensing; and the paper \cite{NikoloskaConformal2025} introduced online design strategies for variational quantum sensing. 

\noindent \textbf{Main Contributions:}  Inspired by these advances, this paper proposes a variational QISAC protocol that adaptively learns the joint measurement strategy and classical processing needed to balance sensing and communication  performance. As illustrated in Fig. \ref{fig:sensing_scheme_indirect}, in the proposed framework, transmitter (Alice) and receiver (Bob) run a variant of superdense coding designed to  simultaneously perform sensing and communication . While Alice follows a standard superdense coding strategy, with a possible communication rate back-off, Bob runs a variational quantum circuit to optimize the measurement strategy, along with neural network-based decoders and estimators for classical post-processing. The hybrid classical-quantum system is trained end-to-end to optimize a weighted sum of communication-centric and sensing-centric performance criterion.  We explore qudit-based transmission strategies that back off the transmission rate permitted by superdense coding to facilitate quantum sensing, and we provide numerical results that illustrate this trade-off.


\vspace{-2mm}

\section{Problem Formulation}
We consider the scenario illustrated in Fig. \ref{fig:sensing_scheme_indirect}  in which a transmitter (Alice) sends classical information to a receiver (Bob) over a channel that depends on an unknown parameter. Specifically, Alice wishes to communicate a message $m$, encoding $B$ bits, to Bob, while Bob simultaneously aims to estimate an unknown parameter $x$ that affects the channel between Alice and Bob. The parameter $x$ is assumed to be discrete, taking one of $K$ possible values in the range $[-\pi,\pi]$. The objective is to ensure both reliable message transmission and accurate parameter estimation. 

As shown in Fig.~1, a third party (Charlie) prepares a maximally entangled two-qudit state $|\psi\rangle = \frac{1}{\sqrt{d}}\sum_{k=0}^{d-1}|k,k\rangle$ for two $d$-dimensional qudits, and distributes one qudit to Alice and the other to Bob. This entangled pair serves as a joint {probe state} that facilitates the ISAC operation. 

\vspace{-2mm}

\subsection{Messages, Probes, and Measurements}
Following superdense coding, Alice encodes the message $m$ on her qudit by applying a unitary $U(m)$ chosen from a set of possible encoding operations. We represent the message as a pair of integers  $m=(m_1,m_2)$, where $m_1,m_2 \in\{ 0,1,\ldots,d'-1\}$ with $d'\in \{1,...,d\}$. Accordingly, the number of transmitted bits is $B=2\log_2(d')$. In conventional superdense coding the number of transmitted bits is $B_\text{max}=2\log_2(d)$, and hence the communication rate back-off applied by Alice as compared to superdense coding is given by the difference    \begin{equation}
\Delta B=B_{\text{max}}-B=2\log_2\left(\frac{d}{d'}\right).
\end{equation} In particular, setting $d'=d$ yields superdense coding, with rate back-off $\Delta B=0$, while smaller values of $d'$ indicate a positive back-off. With $d'=1$, the transmitted rate is $B=0$, and the rate back-off takes the maximum value $\Delta B=B_{\text{max}}$.

Following  superdense coding, the encoding of a message $m=(m_1,m_2)$ is realized via generalized Pauli operators on a $d$-level qudit, i.e., 
\begin{equation}
    U(m) = X^{m_1} Z^{m_2}.
    \label{eq:Um}
\end{equation}
In \eqref{eq:Um}, $X$ and $Z$ are the $d$-dimensional shift and clock operators, which act on computational basis states $|k\rangle$, for $k\in\{0,\dots,d-1\}$, as
\begin{equation}
    X\,|k\rangle = |(k+1)\bmod d\rangle~,
    \label{eq:X}
\end{equation}
\begin{equation}
    Z\,|k\rangle = \omega^k |k\rangle~,  \text{ with } \omega = e^{2\pi i/d}.
    \label{eq:Z}
\end{equation}
Applying the unitary $U(m)$ on Alice's half of the entangled state $|\psi\rangle$  yields the state
\begin{equation}
    |\psi(m)\rangle = (U(m)\otimes I)\,|\psi\rangle,
    \label{eq:psi_m}
\end{equation}
which is shared between Alice (first qudit) and Bob (second qudit).

Alice  sends her qudit through the parameter-dependent channel $U(x)$ to Bob.  Typically, the  unitary channel $U(x)$ takes the form $U(x)=\exp(i\theta G)$ for some Hamiltonian $G$.  After the application of the unitary channel, the joint state of the two qudits becomes 
\begin{equation}
    |\psi(m,x)\rangle = (U(x)\otimes I)\,|\psi(m)\rangle~,
    \label{eq:psi_mx}
\end{equation}
which depends on both the message $m$ and the parameter $x$. In this way, the qudit sent by Alice serves a dual role: it carries the encoded message and simultaneously acts as a probe of the parameter $x$.

Upon receiving Alice's qudit, Bob holds two qudits: his original entangled qudit from Charlie, and Alice's qudit that has passed through the channel. Bob's task is to perform a joint measurement on the two-qudit system to extract information about both message $m$ and parameter $x$. To this end, Bob applies a decoding quantum circuit, parameterized by a set of adjustable angles $\mu$, before measuring. The decoding circuit is designed to entangle and rotate the two qudits into a basis that is informative for both the sensing and communication  tasks.

In our design, the decoding circuit comprises a fixed part followed by a tunable part. Specifically, Bob first applies the two-qudit gate sequence $(H\otimes I)\,\mathrm{CNOT}$, where $H$ is the Hadamard (Fourier) gate acting on Alice's qudit and $\mathrm{CNOT}$ is a controlled-NOT (increment) gate with Alice's qudit as control and Bob's as target. These operations constitute the standard Bell measurement transform used in superdense coding \cite{Harrow2004}. Next, Bob applies a local two-qudit unitary $U(\mu)$, which is a parameterized rotation depending on the vector of variational parameters $\mu$. The role of the unitary  $U(\mu)$ is to define a rotated measurement basis that accounts for the sensing task. 

The overall decoding operation produces the pre-measurement state 
\begin{equation}
    |\psi_{\mu}(m,x)\rangle = U(\mu)\,(H\otimes I)\,\mathrm{CNOT}\;|\psi(m,x)\rangle.
    \label{eq:psi_mu}
\end{equation}
 Note that if parameter $\mu$ is set such that the decoding unitary is an identity transformation, i.e., $U(\mu)=I$, then this transformtion recovers the standard superdense coding protocol \cite{Harrow2004}. 

Bob then performs a projective measurement on the two-qudit state $|\psi_{\mu}(m,x)\rangle$, yielding an outcome $s\in\{0,1,...,d-1\}\times\{0,1,...,d-1\} $. Let $\{|s\rangle\}$ denote the computational basis on the two-qudit Hilbert space with respect to which the measurement is made. The classical outcome $s$ occurs with probability given by Born's rule 
\begin{equation}
    p_{\mu}(s|x) = \big|\langle s \mid \psi_{\mu}(m,x)\rangle\big|^2~,
    \label{eq:likelihood}
\end{equation}
which generally depends on $m$ and $x$. The outcome $s$ thus contains information about both the transmitted message $m$ and the parameter of interest $\mu$.

After the measurement, as shown in Fig. \ref{fig:sensing_scheme_indirect}, Bob processes the classical output $s$ using two classical functions: a \emph{decoder} that produces an estimate $\hat{m}$ of the transmitted message, and an \emph{estimator} that produces an estimate $\hat{x}$ of the unknown parameter. We denote the decoder function by $\hat{m}=f_{\theta}(s)$ and the estimator function by $\hat{x}=g_{\phi}(s)$, where $\theta$ and $\phi$ are the adjustable parameters of the decoder and estimator, respectively. In our implementation, $f_{\theta}$ and $g_{\phi}$ are realized by classical  neural networks that take the measurement outcome $s$ as input and output probability distributions over possible messages or parameter values, as further described in Sec.~III. Bob's final decisions $\hat{m}$ and $\hat{x}$ are the values that maximize these posterior probabilities.

\vspace{-2mm}

\subsection{Design Goal}
Our design goal is to jointly optimize communication reliability and sensing accuracy. Let $P_{\text{succ}}(\mu,\theta, \phi)=\Pr\{\hat{m}=m \}$ denote the probability of successfully decoding the message, and let $P_{\text{acc}}(\mu,\theta, \phi)=\Pr\{\hat{x}=x \}$ denote the probability of accurately estimating the parameter. These probabilities depend on the entire system design, encompassing quantum encoding, quantum decoding, and classical processing. While we ideally seek to maximize both probabilities  $P_{\text{succ}}(\mu,\theta, \phi)$ and $P_{\text{acc}}(\mu,\theta, \phi)$ simultaneously, in practice, there is a trade-off: improving one often comes at the expense of the other. 

To handle this constraint, we formulate the weighted optimization problem
\begin{equation}
\max_{\mu,\;\theta,\;\phi} \;\; \mathbb{E}_{x,m}\Big[w_\text{acc}P_{\text{acc}}(\mu,\theta, \phi) + w_\text{succ}P_{\text{succ}}(\mu,\theta, \phi)\Big]~,
    \label{eq:objective}
\end{equation}
where the expectation is over the distribution of the random parameter $x$ and message $m$, and $w_\text{succ}, w_\text{acc} \ge 0$ are scalar weighting factors that set the relative priority of sensing accuracy versus communication reliability. 

\vspace{-2mm}

\section{Quantum Integrated Sensing and Communication via Variational Optimization}
We address problem \eqref{eq:objective} using a  variational optimization approach, alternating between classical and quantum parameter updates. The receiver design $(\mu, \theta, \phi)$ is optimized iteratively: for a fixed quantum measurement parameter $\mu$, the classical decoder parameters $\theta$ and estimator parameters $\phi$ are trained to maximize the likelihood of correct decisions; and for fixed $\theta,\phi$, the quantum circuit parameters $\mu$ are updated  to improve the weighted objective. 



\vspace{-2mm}

\subsection{Decoding and Estimation via Deep Neural Networks}
We implement the classical decoder $f_{\theta}(s)$ and estimator $g_{\phi}(s)$ as feed-forward  neural networks trained on data collected from the quantum measurement. The decoder network $f_{\theta}(s)$ takes the measurement outcome $s$ as input, and outputs a probability distribution $f_{\theta}(s)=\{p_{\theta}(\hat{m}|s)\}_{\hat{m}}$ over the $(d')^2$ possible messages. We use a standard softmax output layer, yielding 
\begin{equation}
    p_{\theta}(\hat{m}|s) = \frac{\exp(z_{\hat{m}})}{\sum_{\hat{m}'} \exp(z_{\hat{m}'})},
    \label{eq:decoder_softmax}
\end{equation}
where $z_{\hat{m}}$ is the logit corresponding to message hypothesis $\hat{m}$ and the sum runs over all $(d')^2$ messages. The decoded message is then chosen as the most likely message:
\begin{equation}
    \hat{m} = \arg\max_{\hat{m}'}\; p_{\theta}(\hat{m}' | s)~.
    \label{eq:decoder_decision}
\end{equation}

 In a similar manner, the estimator network $g_{\phi}(s)$ outputs a distribution $g_{\phi}(s)=\{p_{\phi}(\hat{x}|s)\}_{\hat{x}}$ over the $K$ possible parameter values $\hat{x}$, and  the parameter estimate is 
\begin{equation}
    \hat{x} = \arg\max_{{\hat{x}'}}\; p_{\phi}(\hat{x}'| s)~.
    \label{eq:estimator_decision}
\end{equation}

\begin{figure*}[htbp]
    \centering
    \subfigure{\includegraphics[width=0.41\textwidth]{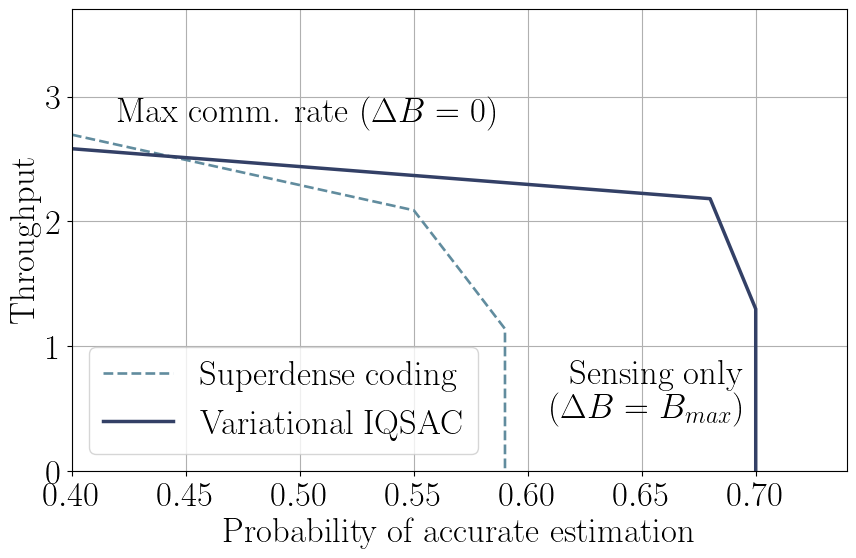}}
    \subfigure{\includegraphics[width=0.41\textwidth]{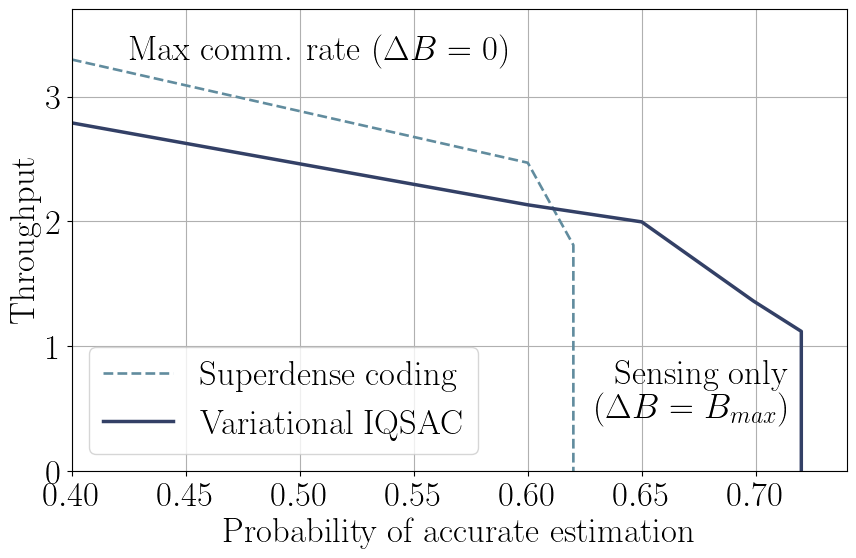}}
    \caption{Communication throughput (bits per channel use) versus  probability of accurate sensing estimation for $d=8$ (left) and $d=10$ (right). The curves are obtained by varying the communication back-off from $\Delta B=0$ to $\Delta B= B_{\max}$. }
    
    \label{fig:entropy_Tr}
\end{figure*}


\vspace{-5mm}

\subsection{End-to-End Variational Optimization}
During training, we optimize the parameters $\theta$, $\phi$, and $\mu$ in an alternating fashion to maximize the objective \eqref{eq:objective}. For a fixed quantum measurement parameter $\mu$,  we train the decoder network $f_{\theta}(s)$ by minimizing the standard surrogate loss given by the cross-entropy loss.  This is achieved using a set of training samples $(s, m)$ obtained from simulated transmissions. The decoder update then takes the gradient descent form
\begin{equation}
    \theta \leftarrow \theta - \eta_{\theta}\, \nabla_{\theta}\Big[-\sum_{(s,m)}  \log p_{\theta}(m | s)\Big]~,
    \label{eq:decoder_update}
\end{equation}
where the sum is over a batch of data points $(s,m)$.  Similarly, for a fixed measurement and decoder, we train the estimator network $g_\phi(s)$ by minimizing the cross-entropy surrogate for parameter classification 
\begin{equation}
    \phi \leftarrow \phi - \eta_{\phi}\, \nabla_{\phi}\Big[-\sum_{(s,x)} \log p_{\phi}(x | s)\Big]~,
    \label{eq:estimator_update}
\end{equation}
with $(s,x)$ representing a data point and $\eta_{\phi}$ being the learning rate.

Finally, for fixed parameters $\theta$ and $\phi$, we optimize the quantum circuit parameters $\mu$. We employ standard gradient ascent  via the parameter-shift rule \cite{schuld2021machine,simeone2022introduction}, approximating the update
\begin{align}
    \mu \leftarrow \mu + \eta_{\mu}\, &\nabla_{\mu}\Big[w_x \E_{p_{\mu}(s|x)} \left[\log p_{\phi}(x|s) \right] \nonumber\\
    &+ w_m \E_{p_{\mu}(s|x)} \left[\log p_{\theta}(m|s)\right]\Big]~,
    \label{eq:mu_update}
\end{align}
where $\eta_{\mu}$ is the learning rate. 


\vspace{-2mm}

\section{Experiments}
\vspace{-2mm}
\subsection{Setup}
We evaluate the proposed variational QISAC protocol via numerical simulation. We consider $d$-level qudit systems with dimensions  $d=8$ and $d=10$, and we examine different values of the rate back-off $\Delta B$.  We model the channel-dependent unitary as a phase rotation on Alice's qudit $U(x) = R_Z (x)$, acting as
\begin{align}
    R_Z (x) |k\rangle = \exp\left(-i \frac{x}{d}\omega^k \right) |k\rangle,
\end{align}
with $\omega^k$ given by \eqref{eq:Z}. The unknown parameter $x$ takes $K=4$ discrete values uniformly spaced in $[-\pi,\pi]$.  This simple channel model captures the essential feature that the parameter modulates the relative phase of the quantum state.

\vspace{-2mm}
\subsection{Implementation}
The decoder and estimator are implemented as fully-connected feedforward neural networks. Each network has two hidden layers with 1024 neurons each and ReLU activation functions. The decoder's output layer has $(d')^2$ neurons,  one per possible message, and the estimator's output layer has $K$ neurons, one per possible parameter value. Both networks are trained using the Adam optimizer with learning rates $\eta_{\theta}=\eta_{\phi}=0.001$. In each outer iteration of the variational training, we perform 100 gradient descent steps on the decoder and 100 steps on the estimator, using mini-batches of simulated training samples.  We perform 100 gradient ascent steps for parameters $\mu$ per iteration, with learning rate $\eta_{\mu}=0.01$. We set $w_x=1$ and $w_m=1$. The entire training loop runs for $T=10$ outer iterations.

\vspace{-2mm}
\subsection{Results}
We quantify communication performance by the throughput and sensing performance by the estimation accuracy. The throughput, measured in bits per channel use, is defined as the product $B P_{\text{succ}}$, i.e., as the number of transmitted bits $B$ times the success probability $P_{\text{succ}}$. Estimation accuracy is measured as the fraction of correctly estimated parameter values.

Fig.~2 plots the trade-off between throughput and accuracy as the communication rate back-off ranges from $\Delta B= 0$, corresponding to the maximum communication rate, to $\Delta B= B_{\max}$, in which case all resources are devoted to sensing. We observe that increasing the communication back-off $\Delta B$ improves the sensing accuracy   $P_{\text{acc}}$ at the cost of a reduced throughput, illustrating the fundamental trade-off between sensing and communication. 

We also compare the variationally optimized measurement with a conventional superdense-coding  measurement. This approach corresponds to setting parameters $\mu$ so that the unitary $U(\mu)$ equals the identity matrix $I$. Our results indicate that the optimized measurement provides a significant increase in the sensing performance $P_{\text{acc}}$. Notably, for intermediate values of the back-off rate $\Delta B$, the variational protocol achieves comparable throughput, while maintaining high estimation accuracy.

\vspace{-2mm}

\section{Discussion and Conclusion}
We have presented a variational protocol for joint quantum communication and sensing, demonstrating how a quantum receiver can be trained to balance the dual tasks of decoding classical messages and estimating unknown parameters. The approach leverages entangled probe states and hybrid quantum-classical optimization to adapt the receiver's measurement strategy. Through simulations, we characterized the trade-off between communication throughput and sensing accuracy and showed that our QISAC design can operate in a regime of simultaneous non-zero rate and high sensing precision. 

The proposed framework is flexible and can be extended in several directions. While we focused on discrete parameter estimation, continuous-valued parameters or multi-parameter sensing scenarios can be considered. Similarly, more expressive parameterized circuits $U(\mu)$ could further improve performance at the expense of circuit complexity. Techniques for robust training, such as advanced optimizers or regularization to avoid barren plateaus, are also worth exploring. Ensuring reliable performance under noise and model uncertainty is another important direction: for instance, recent work has studied online calibration using conformal inference to control estimation loss in variational quantum sensing \cite{NikoloskaConformal2025}, and integrating such techniques could make QISAC receivers more resilient to real-world imperfections.

\bibliographystyle{IEEEtran}
\bibliography{litdab.bib}

\end{document}